\documentstyle[preprint,aps]{revtex}

\newcommand{\AdS}{$\mbox{AdS}_{d+1}\,\,\,$}
\newcommand{\be}{\begin{equation}}
\newcommand{\ee}{\end{equation}}
\newcommand{\bea}{\begin{eqnarray}}
\newcommand{\eea}{\end{eqnarray}}
\newcommand{\nn}{\nonumber \\}

\begin{document}
\tightenlines


\preprint{}

\title{QUANTIZATION IN ADS AND THE ADS/CFT CORRESPONDENCE}

\author{Victor O. Rivelles}

\address{Center for Theoretical Physics \\ Massachusetts Institute of
  Technology \\ Cambridge, MA  02139, USA \\ and \\Instituto de
  F\'{\i}sica \\ Universidade de S\~{a}o Paulo\\  Caixa Postal 66318,
  05315-970, S\~{a}o Paulo, SP, Brazil\\ E-mail: rivelles@lns.mit.edu}

\maketitle


\begin{abstract}The quantization of a scalar field in AdS leads to two kinds
of normalizable modes, usually called regular and irregular modes. The
regular one is easily taken into account in the standard prescription
for the AdS/CFT correspondence. The irregular mode requires a
modified prescription which we argue is not completely
satisfactory. We discuss an alternative quantization in AdS which
incorporates boundary terms in a natural way. Within this quantization
scheme we present an improved prescription for the AdS/CFT 
correspondence which can be applied to both, regular and irregular
modes. Boundary conditions other than Dirichlet are naturally treated
in this new improved setting.

\keywords{AdS/CFT Correspondence, Anti DeSitter Space-Time, Quantum
  Fields in Curved Space-Time}

\end{abstract}
 
\newpage

\section{Introduction}

The AdS/CFT correspondence relates a quantum field theory in the bulk
of AdS with a conformal field theory living at its
border\cite{Witten:1998qj}. In the simplest case, that for a scalar  
field, the quantization 
in the bulk produces two kinds of normalizable modes depending on the
mass of the field\cite{Breitenlohner:jf}. For large masses there is
only one mode, the regular one, while for small masses regular and
irregular modes can propagate in the bulk. The AdS/CFT correspondence,
as originally proposed\cite{Witten:1998qj}, considered only the
conformal theory originated from the 
regular mode. Subsequently, a modified prescription was presented to
take into account the irregular mode as well
\cite{Klebanov:1999tb}. When one of them is mapped to the border the
other one  becomes the source for the conformal operator. 
However, it is not clear, in both prescriptions, which modes are being
taken into account.  It seemed that there was no way for selecting
which mode is mapped and which one is becoming the 
source. Then, it was proposed an alternative quantization in AdS which
can effectively select which mode is propagating in the
bulk\cite{Minces:2001zy}. Within  
this new framework the AdS/CFT correspondence can be improved in the
sense that it is possible to keep track of the modes in the bulk. In
this new situation we have a precise picture of the physics in the bulk
and in the border. 

In Section 2 we present the usual quantization in AdS. In the
next section we present the AdS/CFT correspondence for regular and
irregular modes and discuss its limitations. In Section 4 we
present the alternative quantization in AdS. Finally, in Section 5 we
show how the AdS/CFT prescription can be improved.  

\section{Quantization in \AdS}

Consider a scalar field $\phi$ with mass $m$ in AdS. Its energy is
defined in the standard way through the canonical energy-momentum
tensor $T^{\mu\nu}$. The quantization of
this field, with conserved, positive and finite energy, and no energy
flux at the border, leads to the regular mode $\phi_R$. Near the
border, at $\rho =\pi/2$, it behaves as $\phi_R = \epsilon^{\Delta_+}$, 
where $\epsilon = \cos\rho$ and 
\be
\Delta_\pm = \frac{d}{2} \pm \nu, \qquad \nu = \sqrt{\frac{d^2}{4} +
  m^2}.
\ee

Since the energy-momentum tensor is defined up to improvement terms 
\be
t_{\mu\nu} = T_{\mu\nu} + \beta ( g_{\mu\nu} D^2 - D_\mu D_\nu +
R_{\mu\nu} ) \phi^2,
\ee
we must look for other solutions whose energy, derived from the
improved energy-momentum tensor, is also conserved, 
positive and finite. It is then found that there is another solution
whose asymptotic behavior is $\phi_I = \epsilon^{\Delta_-}$ if $\beta
= \Delta_-/(2 \Delta_- + 1)$ and the mass is constrained by $ 0 \le
\nu < 1$. This is the so called irregular mode.  
For any value of the mass, stability of the solutions also requires
$\nu$ real which is known as the Breitenlohner-Freedman bound $
m^2 \ge - d^2/4$\cite{Breitenlohner:jf}.  

We then have the following picture of the physics in the bulk. For
$\nu>1$ only the regular mode can be quantized and the irregular
one is a classical background. For $0 \le \nu < 1$ both modes can be
quantized and we have two quantum theories in the bulk. 

\section{The AdS/CFT Correspondence}

In the usual prescription for the AdS/CFT
correspondence\cite{Witten:1998qj} we compute   
the partition function with the fields $\phi$  having fixed values
$\phi_0$ at the border of AdS. These asymptotic fields are then
regarded as sources for the conformal theory operator $\cal O$ which
lives in the AdS border 
\be
\label{AdS/CFT0}
\int_{\phi_0} {\cal D} \phi \, e^{-S[\phi]} = < e^{\int d^dx\,\, {\cal O}
  \phi_0} >.
\ee
This proposal has been verified in a number of
  situations\cite{Aharony:1999ti}. In order to compute the lhs of
  (\ref{AdS/CFT0}) we consider, in first 
approximation, only the classical solution so that we have 
\be
\label{AdS/CFT}
e^{-S[\phi_0]} = < e^{\int d^dx\,\, {\cal O}  \phi_0} >.
\ee

To proceed, it is convenient to use the Euclidean version of AdS and
also Poincar\'e coordinates  
\be
ds^2 = \frac{1}{x_0^2} \sum_\mu ( dx^\mu )^2,
\ee
so that the border is now at $x_0 = 0$. Since the metric diverges at
the border we must be careful in defining how we approach
it\cite{Freedman:1998tz}. Near the border the fields will be
denoted by $\phi_\epsilon (\vec{x}) = \phi(\epsilon, \vec{x})$. We
then absorb all divergences in $\phi_\epsilon (\vec{x})$ before
taking the limit $\epsilon \rightarrow 0$. For Dirichlet boundary
conditions we find, after using the equation of motion 
\be
\label{ac1}
S = - \frac{1}{2} \int d^{d+1}x \,\, \sqrt{g} \,\, (\partial\phi \partial
\phi + m^2 \phi^2) = \frac{1}{2} \int d^d x \,\, \sqrt{h} \,\, 
\phi_\epsilon \partial_n \phi_\epsilon,
\ee
where $h$ is induced metric at the border and $\partial_n$ is the
normal derivative. Then, the action is always reduced to a boundary
term when it is on-shell. Inserting the solution of the Klein-Gordon
equation  we get 
\bea
\label{expansion1}
S &=& \int d^dx \, d^dy \,\, \phi_\epsilon(\vec{x})
\phi_\epsilon(\vec{y}) \times \nn &&\int d^d k \,\, e^{-i\vec{k}\cdot
  (\vec{x}-\vec{y}) } \epsilon^{-d} \left[ \Delta_- +
  \frac{(k \epsilon)^2}{2(1-\nu)}  -
  \frac{\Gamma(1-\nu)}{ 2^{2\nu-1} \Gamma(\nu)} (k\epsilon)^{2\nu} +
  \dots \right],
\eea
where $\dots$ means high order terms in $\epsilon$. The first two terms
are proportional to a delta function and its second derivative,
respectively. They 
give rise to contact terms which are usually disregarded since we are
interested only in the terms which do not vanish when $\vec{x} \not=
\vec{y}$. Also, the leading term gives $\epsilon^{2\nu-d} =
\epsilon^{-2\Delta_-}$ which means that to remove all divergences
\be
\label{limit1}
\phi_\epsilon = \lim_{\epsilon \rightarrow 0} \epsilon^{\Delta_-}
\phi_0.
\ee
When the integration in $\vec{k}$ is performed we get 
\be
\label{delta+}
S = \int d^dx \, d^dy \,\, \frac{\phi_0(\vec{x})
  \phi_0(\vec{y})}{|\vec{x}-\vec{y}|^{2\Delta_+}}.
\ee
After using (\ref{AdS/CFT}) we find that the conformal dimension of
$\cal O$ is $\Delta_+$ so that the regular mode was captured by the
correspondence. From (\ref{limit1}) we learn that the source for it, 
$\phi_\epsilon$, behaves near the border as $\epsilon^{\Delta_-}$ and
that is the irregular mode. So, even though no restriction has been
  assumed for $\nu$ we are in the situation $\nu>1$ since only the regular
mode was quantized in the bulk. 

To capture the irregular mode Klebanov and
Witten\cite{Klebanov:1999tb} proposed that 
instead of using the action (\ref{ac1}) in (\ref{AdS/CFT0}) we should
use its Legendre transform but not for the full action (\ref{ac1}),
only for its leading terms in (\ref{expansion1}). In this situation,
the Legendre transform is then defined as 
\be
\label{Legendre1}
\tilde{S}[\tilde{\phi}_0] = S[\phi_0] + (2\Delta_- -d) \int
\frac{d^dk}{(2\pi)^d} \phi_0(\vec{k})  \tilde{\phi}_0(-\vec{k}),
\ee
and we get 
\be
\tilde{S}[\tilde{\phi}_0] = \int d^dx \, d^dy \,\,
\frac{\tilde{\phi}_0(\vec{x}) 
  \tilde{\phi}_0(\vec{y})}{|\vec{x}-\vec{y}|^{2\Delta_-}}.
\ee
Now, using (\ref{AdS/CFT}) for $\tilde{S}$ we get that the conformal
dimension of $\cal O$ is indeed $\Delta_-$, so that the irregular mode
was taking into account. 
This means that $\nu$ is in the range $0\le\nu<1$ since this is the
condition for the irregular mode to be quantized in the bulk. 
Notice, however, that no assumption on $\nu$ was made. So, in both
cases, the physical picture is quite obscure since we need to impose
by hand which mode is propagating in the bulk and which mode remains 
classical. 

Moreover, the Legendre transformation (\ref{Legendre1}) can be carried
out  for any value of $\nu$, and not only for $0\le\nu<1$, since no
restriction on $\nu$ was assumed. 
Also this prescription does not work when $\nu=0$, that is, when
$\Delta_+ = \Delta_-$. In this case we would expect that the action
and its Legendre transform would be proportional to each
other. However, the leading term in
(\ref{expansion1}) is now $\log k$ and its Legendre transform is not
proportional to itself. So we will look for an alternative formulation
where we can have more control on the relevant modes of the scalar
field. 

\section{Alternative Quantization in \AdS}

As discussed in the previous section, only boundary terms contribute to
the correspondence. So we would like to consider a quantization
framework were they are naturally taken into
account\cite{Minces:2001zy}. In the usual  
quantization, the energy is defined through the energy-momentum
tensor. In its computation, boundary terms are disregarded. In fact,
the addition of boundary terms to the action corresponds to improving
the energy momentum-tensor. A definition of energy, which takes into
account all boundary terms, is the one which makes use of the Noether
current for time displacements\cite{Minces:2001zy}. If we have a
conserved Noether current 
$J^\mu$ and a Killing vector for infinitesimal displacements,
$\xi^\mu$, we can define the energy as
\be
E = \int d^dx \,\, \sqrt{g} J^\mu \xi_\mu.
\ee
Conservation of energy implies that 
\be
\dot{E} = \int d^dx \,\, \sqrt{g} \, [ - D_i (J^i \xi_0 ) + \xi_i D_0 J^i
],
\ee
so that it is a total derivative for time translations. 
Again, energy is conserved if there is no flux at
the border and this will impose conditions on the solutions of the
Klein-Gordon equation\cite{Minces:2001zy}. 

An important fact about AdS has to do with the mass term. Since the
AdS curvature is constant the origin of the quadratic term in $\phi$
is ambiguous. It can have a piece coming from the quadratic term in the
potential but can also have another part coming from a non-minimal
coupling to the background. We then have to consider the action 
\be
\label{action2}
S_0 = - \frac{1}{2} \int d^dx \,\, \sqrt{g} \, [ \partial\phi \partial\phi
  + (m^2 + \lambda R) \phi^2 ],
\ee
where $\lambda$ is the non-minimal coupling constant and $R=-d(d+1)$ is
the Ricci scalar. This means that $\phi$ has an effective mass $M^2 =
m^2 + \lambda R$. This may seems innocuous but the two quadratic terms
couple differently to gravity so that when the metric is varied they
give different contributions as we shall see immediately.

When we perform infinitesimal variations of the metric and require
that the variation of the action vanishes we find a boundary term,
reminiscent from the Gibbons-Hawking term. Then we must add to
(\ref{action2}) a boundary contribution
\be
\label{action3}
S = S_0 + \lambda \int d^dx \,\, \sqrt{h} K \phi^2,
\ee
where $K$ is the trace of the extrinsic curvature on the boundary
which, in global coordinates, is given by $K = -
(d-\cos^2\rho)/\sin\rho$. 

Performing now infinitesimal variations of $\phi$ and using the field
equations we find
\be
\delta S = - \int d^dx \,\, \sqrt{h} ( \partial_n \phi - 2 \lambda K
\phi) \delta \phi.
\ee 
For Dirichlet boundary condition this term vanishes but for other
boundary conditions it does not. We must then add further boundary
terms to the action (\ref{action3}). This has been discussed in
detail\cite{Minces:2001zy,Minces:1999eg} but here we will 
consider only the Dirichlet case for 
simplicity. Then, no further boundary terms are required by the
variational principle. 

The Noether current for infinitesimal translations can now be computed
and we find
\be
J^\mu = - T^\mu_0 - \lambda [ \delta^\mu_0 D_\nu ( K n^\nu \phi^2 ) -
  K n^\mu \partial_0 \phi^2 ],
\ee
where $T^{\mu\nu}$ has now the effective mass $M$
instead of $m$.  We have then improved the energy-momentum tensor in a
natural way using the Noether current. The energy is them simply 
$E = \int d^dx \,\, \sqrt{g} J^0$. 

The quantization now proceeds along familiar lines. Requiring that
the energy be conserved, positive and finite leads to the two
modes with asymptotic behavior $ \phi_R = \epsilon^{\Delta_+}$ and $
\phi_I = \epsilon^{\Delta_-}$, where now
\be
\Delta_\pm = \frac{d}{2} \pm \nu, \qquad \nu = \sqrt{\frac{d^2}{4} +
  M^2}.
\ee
For the regular mode we find no constraints. For the irregular one we
again find that $0\le\nu<1$ and $ \lambda = \Delta_-/2d$.
There are two solutions for this equation $\lambda = [d-1 \pm
  \sqrt{(d-1)^2+16m^2}]/(8d)$. We also found that the
  Breitenlohner-Freedman bound is still required but now with $m$
  replaced by the effective mass $M$. 

This same analysis can be performed for other boundary conditions with
similar conclusions\cite{Minces:2001zy}. 

\section{Improved AdS/CFT Prescription}

As discussed before we will consider an improved AdS/CFT
prescription which holds for any boundary condition. It reads
\be
e^{-S[A_0]} = < e^{ \int d^dx \,\, {\cal O} A_0} >,
\ee
where $A_0$ is the combination of fields which is fixed at the
boundary. For the Dirichlet case it is simply $\phi_0$. 

We now compute the action (\ref{action3}) for the solution of the
Klein-Gordon equation with Dirichlet boundary condition
\be
S = \frac{1}{2} \int d^dx \,\, \sqrt{h} \phi_\epsilon ( \partial_n
\phi_\epsilon - 2 \lambda K_\epsilon \phi_\epsilon ).
\ee
Since there is a boundary term in (\ref{action3}) we now find that 
\bea
\label{action4}
&S& = \int d^dx \, d^dy \,\, \phi_\epsilon(\vec{x})
\phi_\epsilon(\vec{y}) \times \nn  && \int d^d k \,\, e^{-i\vec{k}\cdot
  (\vec{x}-\vec{y}) } \epsilon^{-d} \left[ \Delta_- - 2\lambda d + 
  \frac{(k \epsilon)^2}{2(1-\nu)}  -
  \frac{\Gamma(1-\nu)}{ 2^{2\nu-1} \Gamma(\nu)} (k\epsilon)^{2\nu} +
  \dots \right].
\eea
As for the Legendre transformed action we now consider the full action
(\ref{action4}) and not just its leading term. We then find
\bea
\label{action5}
&\tilde{S}& = \int d^dx \, d^dy \,\, \tilde{\phi}_\epsilon(\vec{x})
\tilde{\phi}_\epsilon(\vec{y}) \times \nn && \int d^d k \,\,
e^{-i\vec{k}\cdot 
  (\vec{x}-\vec{y}) } \epsilon^{-d} \left[ \Delta_- - 2\lambda d + 
  \frac{(k \epsilon)^2}{2(1-\nu)}  -
  \frac{\Gamma(1-\nu)}{ 2^{2\nu-1} \Gamma(\nu)} (k\epsilon)^{2\nu} +
  \dots \right]^{-1}.
\eea

Let us consider the case where only regular modes propagate, that
is when $\lambda \not= \Delta_- /(2d)$. From (\ref{action4}) we find
the same result as in the usual case (\ref{delta+}). For the Legendre
transformed action
$\tilde{S}$ we have to invert the term in square
brackets and when that is done it reproduces the same $\epsilon$
structure as that in (\ref{action4}). We then get $\tilde{S} =
1/(\Delta_- - 2\lambda d)^2 S$ so that it is also capturing the
regular mode in AdS. So both, the action an its Legendre transform, 
give a boundary theory with conformal dimension $\Delta_+$. This is
consistent with the condition $\lambda \not= \Delta_- /(2d)$ which
does not allow irregular modes to be quantized in the bulk. We can
further check that 
\be 
\phi_\epsilon = \epsilon^{\Delta_-} \phi_0, \qquad
\tilde{\phi}_\epsilon = \epsilon^{\Delta_-} \tilde{\phi}_0,
\ee
showing that the irregular modes are the classical sources for $\cal
O$. 

Now we can consider the case when regular and irregular modes
propagate in the bulk, that is, $\lambda = \Delta_- /(2d)$. Note that
now there are no contact terms in (\ref{action4}). From the action
(\ref{action4}) we find that the conformal dimension is $\Delta_+$, so
that it is associated with the regular mode propagation in the
bulk. Now the Legendre transformed action (\ref{action5}) has a
different expansion and we get another $\epsilon$ structure. We
then find that the conformal dimension is $\Delta_-$, so that it
is capturing the irregular mode in the bulk. We can also check
that 
\be 
\phi_\epsilon = \epsilon^{\Delta_-} \phi_0, \qquad
\tilde{\phi}_\epsilon = \epsilon^{\Delta_+} \tilde{\phi}_0, 
\ee
in agreement with the expected result. It is worth to remark that the
absence of contact terms in this case implies that no counterterms are
required to remove them. 

We can also consider the case with $\nu=0$ without problems. We find
that with this prescription the action and its Legendre transform are
indeed proportional and the fields have the correct asymptotic
behavior. 

\section{Conclusions}

The natural ambiguity for the mass term in the bulk of AdS required
that we distinguished between contributions from the potential and from a
non-minimal coupling to gravity. The quantization, where use is made of
the Killing vector for time translation to define the energy, leads to
constraints on the non-minimal coupling. These constraints are
associated to the modes which are able to propagate in the bulk. In
the AdS/CFT correspondence these constraints tell us which mode is
propagating in the bulk so that we have a precise picture of which modes are
classical and which ones are quantum. We also improved the
correspondence applying the Legendre transform of the full action and
not just to its leading 
terms. In this way a consistent picture emerges where all values of
the mass and all boundary conditions can be considered. 

\section{Acknowledgments}

I would like to thank P. Minces for the collaborations and
discussions on this subject. This work was partially
supported by CAPES and PRONEX under contract CNPq 66.2002/1998-99.



\end{document}